\documentclass[useAMS,usenatbib,aas_macros]{mn2e} 
\usepackage {graphicx,times,aas_macros}

\newcommand{\equ}[1]{eq.~(\ref{eq:#1})}

\newcommand{\se}[1]{\S\ref{sec:#1}}

\newcommand{\Fig}[1]{Figure~\ref{fig:#1}}
\newcommand{\Figs}[1]{Figures~\ref{fig:#1}}
\newcommand{\tab}[1]{Table~\ref{tab:#1}}
\newcommand{\be}{\begin{equation}}
\newcommand{\ee}{\end{equation}}
\newcommand{\bea}{\begin{eqnarray}}
\newcommand{\eea}{\end{eqnarray}}

\newcommand{\msun}{{\rm M}_\odot}
\newcommand{\Msun}{{\rm M}_\odot}

\newcommand{\ifm}[1]{\relax\ifmmode#1\else$\mathsurround=0pt #1$\fi}
\newcommand{\kms}{\ifmmode\,{\rm km}\,{\rm s}^{-1}\else km$\,$s$^{-1}$\fi}

\newcommand{\kpc}{\,{\rm kpc}}
\newcommand{\pc}{\,{\rm pc}}

\newcommand{\Myr}{\,{\rm Myr}}

\newcommand{\ltsima}{$\; \buildrel < \over \sim \;$}
\newcommand{\lsim}{\lower.5ex\hbox{\ltsima}}
\newcommand{\gtsima}{$\; \buildrel > \over \sim \;$}
\newcommand{\gsim}{\lower.5ex\hbox{\gtsima}}

\def\omm{\Omega_{\rm m}}
\def\oml{\Omega_{\Lambda}}
\def\omb{\Omega_{\rm b}}
\def\sy{\,{\rm M}_\odot\, {\rm yr}^{-1}}

\def\cmc{\,{\rm cm}^{-3}}
\def\cms{\,{\rm cm}^{-2}}

\def\Mv{M_{\rm v}}
\def\Rv{R_{\rm v}}

\def\Ms{M_*}

\def\Ams{\.{A}}
\def\Teq{T$_{\rm eq}$}

\usepackage{color}

\begin{document}

\large

\title[Radiative Feedback]{Radiative feedback and the low efficiency
  of galaxy formation in low-mass haloes at high redshift}

\author[Ceverino et al.]
{\parbox[t]{\textwidth} 
{
Daniel Ceverino$^1$\thanks{E-mail: daniel.ceverino@uam.es}, 
Anatoly Klypin$^2$, 
Elizabeth S. Klimek$^2$, 
Sebastian Trujillo-Gomez$^2$, 
\\ 
Christopher W. Churchill$^2$,
Joel Primack$^3$,
Avishai Dekel$^4$ } \\ \\
$^1$Departamento de F{\'i}sica Te{\'o}rica, Universidad Aut{\'o}noma de Madrid, 28049 Madrid, Spain \\
$^2$Astronomy Department, New Mexico State University, Las Cruces, NM, USA \\
$^3$Department of Physics, University of California, Santa Cruz, CA, USA \\
$^4$Racah Institute of Physics, The Hebrew University, Jerusalem 91904, Israel }

\date{}

\pagerange{\pageref{firstpage}--\pageref{lastpage}} \pubyear{0000}

\maketitle

\label{firstpage}

\begin{abstract}
Any successful model of galaxy formation needs to explain the low rate of star formation in the small progenitors of today's galaxies. This inefficiency is necessary for reproducing the low stellar-to-virial mass fractions, suggested by current abundance matching models.  A possible driver of this low efficiency is the radiation pressure exerted by ionizing photons from massive stars. The effect of radiation pressure in cosmological, zoom-in galaxy formation simulations is modeled as a non-thermal pressure that acts only in dense and optically thick star-forming regions. We also include photoionization and photoheating by massive stars. The full photoionization of hydrogen reduces the radiative cooling in the $10^{4-4.5}$ K regime. The main effect of radiation pressure is to regulate and limit the high values of gas density and the amount of gas available for star formation. This maintains a low star formation rate of $\sim 1 \ \Msun \ {\rm yr}^{-1}$ in halos with masses about $10^{11} \ \Msun$ at $z\simeq3$. Infrared trapping and photoionization/photoheating processes are secondary effects in this mass range. The galaxies residing in these low-mass halos contain only $\sim0.6\%$ of the total virial mass in stars, roughly consistent with abundance matching. Radiative feedback maintains an extended galaxy with a rising circular velocity profile.
\end{abstract}


\section{Introduction}
\label{sec:intro}

The growth of structures on extra-galactic scales is properly captured by the initial conditions predicted by the inflation model and the gravitational processes within the standard $\Lambda$CDM cosmology. However, on galactic scales, baryon physics complicates matters and
introduces major open issues in galaxy formation.

The formation and growth of galaxies is somehow related to the formation and growth of their dark matter halos \citep[][and references therein]{Neistein06, LiMoGao08}. 
However, there is an important difference between the formation of halos and galaxies. 
Halos are built bottom-up, where small halos form first and gradually merge into bigger halos.
On the other hand, observations indicate that massive galaxies formed their stars early while low-mass galaxies formed a large fraction of their stars later, typically showing higher star-formation activity at low redshifts
\citep{Heavens04, ThomasDaniel05, Gallazzi05}.
This implies that star formation was inefficient in the progenitors of the low-mass galaxies at high redshifts.

Achieving this low efficiency of star formation at high redshifts is crucial for any successful galaxy formation model.
It is a necessary condition for matching 
the observed scaling relations and the cosmic SFR history simultaneously \citep{Bouche10}.
It is also necessary for reproducing the downsizing trend in stellar mass.
Otherwise, galaxies with stellar masses between $\Ms=10^9-10^{11} \ \Msun$ form stars too early and are too passive at late times \citep{Fontanot09}.
This is a problem in most semi-analytical models and simulations.
For example, the Eris2 simulation \citep{Shen13} forms too many stars at high redshifts (SFR= $ 20  \sy$ at $z=2.8$). 
As a result, the evolution in the number of galaxies at a fixed stellar mass is too fast at high redshifts, $z>2$, and too slow at lower redshifts, $z<2$, as compared to observations \citep{Weinmann12}.

%
Recent abundance matching techniques \citep{ConroyWechsler09,Moster10, Guo10, Behroozi10, Moster12, Behroozi12}
have constrained the stellar-to-virial mass ratio as a function of the virial mass at different redshifts by making a correspondence between the halo mass function from N-body simulations and the galaxy stellar mass function from different surveys.
This ratio is a strong function of the virial mass and it  decreases strongly at intermediate-to-low mass halos, $\Mv \lsim 10^{11} \ \Msun$, owing to the different shapes of the halo mass function and the galaxy stellar mass function at the low-mass end \citep{BoschXiaohuMo03, Shankar06}. This has been interpreted as a strong evidence of a drop in the efficiency of galaxy formation at low masses \citep{Bosch07,Baldry08,Kravtsov10}.
The models of abundance matching, 
as well as halo occupation distribution modeling \citep{Zheng07, Wake11}
and conditional luminosity function modeling \citep{Yang12},
agree in a low stellar-to-virial mass ratio, f$_{\rm stars}(z=3)=0.06\%-0.6\%$ for halos with masses of $\Mv=10^{11} \ \Msun$ at $z=3$. 
This means that less than 1\% of the total mass within a high-z galactic halo is stellar mass.

%
Supernova feedback from the newly formed stars has been traditionally invoked to regulate or even stop star formation at this mass scale of $\Mv  \lsim  10^{11} \ \Msun$ or $V_{\rm c}  \lsim  100 \kms$ \citep{DekelSilk86}.
However, early simulations of galaxy formation were limited by their low resolution, so feedback was modeled in a very simplistic way \citep{ NavarroBenz91, Katz92, NavarroWhite93, NavarroSteinmetz00,  Abadi03}.
Achieving the right efficiency of supernova feedback has been quite challenging in simulations with low resolution,
in which the thermal energy from supernova explosions is diluted in a large volume, where radiative cooling is very efficient.

%
Several  methods have been proposed to circumvent the problems due to inadequate resolution.
Some of these subgrid models encapsulate the effect of feedback and unresolved multiphase structure as an effective equation of state \citep{SpringelHernquist03, TeyssierChaponBournaud10}.
Other commonly used ad-hoc prescriptions turn off cooling during a period of time right after thermal energy is injected into star forming regions \citep{GerritsenIcke97, ThackerCouchman00, SommerLarsen03, Keres05, Governato07, Agertz11}.
In other models, the feedback energy is stored in an unresolved reservoir of gas, 
which does not cool  \citep{Scannapieco08}.
Other phenomenological models impose a given feedback effect, namely galactic winds  \citep{SpringelHernquist03, DallaVecchiaSchaye08, OppenheimerDave08, Schaye10,VecchiaSchaye12,Vogelsberger13}. 
 
%
There are numerous observational indications that star-forming regions are disrupted by the formation of HII regions inside them.
These HII regions expand due to a combination of photoionization and photoheating processes, as well as radiation pressure by the photons emitted from massive stars in young star clusters \citep{Matzner02, KrumholzMatzner09,Pellegrini11, Walch12}.
The combination of these radiative processes with stellar winds and supernova explosions drive the disruption of giant molecular clouds (GMC) in $\sim30 \Myr$ \citep{Fukui10}, the typical lifetime of a massive star that explodes as a core-collapsed supernova. 
It is unclear which of these mechanisms dominates over the others,
but it is clear that supernova explosions are not the only source of feedback,
since GMCs are affected by radiation before any SNe explode.

The radiation from massive stars is an 
important source of pressure in dense, star-forming regions before supernova explosions \citep{MurrayQuataertThompson10, Hopkins11, Agertz12}.
However, the coupling between gas and radiation depends on the gas optical depth.
If the gas is optically thin ($\tau_{\lambda} << 1$) to radiation of a given wavelength, the radiation escapes freely without exerting much pressure ($P_{\lambda}  \propto [ 1 - \exp(-\tau_{\lambda})] $).
The radiation able to ionize hydrogen shows the strongest coupling with neutral gas. It has a high opacity, $\kappa_{\rm UV}\sim10^3 \cms {\rm g}^{-1}$ 
\citep{ThompsonQuataertMurray05},
and it dominates the total luminosity in star clusters younger than 5-10 Myr.

\cite{MurrayQuataertThompson10} argue that radiation pressure exerted on dust grains is the dominant force that drives the expansion of HII regions in our Galaxy. Dust grains scatter, absorb and re-radiate in all wavelength, so in principle all photons coming from star clusters would generate pressure on a dusty medium,
although the typical flux-mean, dust opacity, $\kappa_{\rm F} = 5 \cms {\rm g}^{-1}$, is much lower than the UV opacity described above.
The effect of radiation pressure could be significantly enhanced if infrared photons are reradiated multiple times.
This multiplicative effect increases the radiation pressure in proportion to the dust optical depth \citep{MurrayQuataertThompson10, Hopkins11, Agertz12}.
However, 
the injection of momentum by radiation on dust grains is poorly understood in environments characterized by high optical depths. \cite{KrumholzThompson12, KrumholzThompson13} argue that 
the multiplicative effect of infrared photons has been overestimated in previous works and
the boost in radiation pressure is much weaker, on the order of unity.

%
Little attention has been given to the feedback effects of photoionization and photoheating from massive stars.
\cite{Cantalupo10} argues that the missing ingredient of feedback could be the effect of photoionization by local sources on the surrounding cold gas. UV and soft X-rays could modify the ionization state of the main gas coolants, such as oxygen.
This effectively decreases the cooling rates by orders of magnitude in warm gas with temperatures between 10$^4$ and 10$^5$ K. This important aspect of radiative feedback remains to be included in most cosmological simulations of galaxy formation.

%
Radiation pressure feedback has just recently been considered in numerical work studying isolated galactic discs \citep{Hopkins11,  Agertz12}. 
\cite{Hopkins11} have proposed a phenomenological implementation of radiation pressure on dust grains that includes the multiplicative effect of the trapped infrared radiation discussed above.
\cite{Agertz12} have developed another 
model that takes into account the momentum injection from stellar winds and supernovae, as well as radiation pressure.
Although the subgrid models are different, they both reach similar conclusions about the importance of radiation pressure in the self-regulation of star formation in idealized examples, from individual star forming clouds to isolated disc galaxies.

%
%
\cite{Brook12} and \cite{Stinson12} have introduced the concept of early feedback.
They assume that 10\% of the bolometric luminosity radiated by young stars is converted to thermal energy before the first supernova explodes. As opposed to previous subgrid models, the temperature of the gas is significantly modified. This overestimates the amount of warm-hot gas with temperatures $10^{5.5}-10^6$ K within the galaxy.
\cite{Brook12} report OVI column densities of 1$0^{15}-10^{16} \cms$ for a $0.4 {\rm L}_*$ galaxy on kpc scales, but typical values of the local ISM are much lower: N(OVI)$=10^{13}-10^{14} \cms$ \citep{Jenkins78,SavageLehner05,Oegerle05}.
Therefore, observations suggest that the extra pressure coming from radiation is a non-thermal component. It provides an extra source of momentum but it is not a source of thermal energy.

%
 \cite{Wise12} have performed radiative-transfer simulations that compute the momentum injection from ionizing photons. These simulations show that the injection of momentum regulates star formation in the first galaxies at the beginning of the reionization epoch.
 It prevents the overcooling regime and runaway star formation in dwarf galaxies inside dark matter halos of $\Mv=2 \times 10^8 \ \Msun$ at $z=8$.
However, the simulations can not proceed to lower redshifts because the ray-tracing algorithm is computationally very expensive after reionization. 
%
Finally, \cite{Aumer13} implement a subgrid model similar to Hopkins' model but in cosmological simulations with a factor of about 100  lower resolution. It remains unclear if the subgrid model proposed in \cite{Hopkins11} can be extrapolated to such large scales without loss of physical meaning
(but see also \cite{Hopkins14}). 

%
This paper is a first step in our modeling of stellar feedback beyond the supernova feedback paradigm.
Our approach is to add more physical processes into simulations of galaxy formation.
In this first paper, we model the effects of ionizing radiation coming from massive and young stars in their surrounding gas. 
These effects include the injection of momentum by radiation pressure, photoionization and photoheating.
The model is local, focused on ~15-30 pc scales, the typical scales of HII regions around 10$^3 \ \msun$ star clusters.
We study the effects of these processes in the regulation of star formation in the high-redshift progenitors of today's galaxies. 
Paper II in this series \citep{Trujillo13} focuses on the assembly of dwarf galaxies and low-mass spirals to $z=0$.
The outline of this paper is the following.
In section \se{models} we describe the model of radiative feedback in detail. 
\se{IC} describes the initial conditions of the test runs.
\se{results} describes the main results of this paper, \se{discussion} is the discussion section and \se{summary} provides the summary.

\section{Modeling of radiative feedback}
\label{sec:models}

The modeling of radiative feedback is done in different steps. Each one adds a different effect of the ionizing radiation on the surrounding gas: radiation pressure, the effects of photoheating and photoionization on the net cooling rates, and the boosting of radiation pressure due to infrared radiation.

\subsection{Base Subgrid Model without Radiative Feedback}
\label{sec:base}

First, we describe the base model, which includes all the standard physics of galaxy formation: cooling, star formation and thermal feedback.
This subgrid model is based on the model first described in \citet[][hereafter CK09]{Ceverino09} and implemented within the \textsc{ART} code \citep{Kravtsov97,Kravtsov03}. 
The \textsc{ART} code 
follows the evolution of a 
gravitating N-body system and the Eulerian gas dynamics using an adaptive mesh 
refinement approach. Beyond gravity and hydrodynamics, the code incorporates 
many of the physical processes relevant for galaxy formation.  
These processes, representing subgrid 
physics, include gas cooling by atomic hydrogen and helium, metal and molecular 
hydrogen cooling, and photoionization heating by a constant cosmological UV background with partial 
self-shielding, star formation and feedback, as described in 
\citet{CDB,Ceverino12}. 

Cooling and heating rates are tabulated for a given gas 
density, temperature, metallicity and UV background based on the \textsc{cloudy} code 
\citep[version 96b4;][]{Ferland98}, assuming a slab of thickness 1 \kpc. 
A uniform UV 
flux based on the redshift-dependent \citet{HaardtMadau96} model is assumed, 
except at gas densities higher than $0.1\cmc$, where a substantially 
suppressed UV background is used
($5.9\times 10^{26} {\rm erg}{\rm s}^{-1}{\rm cm}^{-2}{\rm Hz}^{-1}$) 
in order to mimic the partial self-shielding of dense gas. 
This allows the dense gas to cool down to temperatures of $\sim 300$ K. 
The assumed equation of state is that of an ideal mono-atomic gas. 
Artificial fragmentation on the cell size is prevented by introducing 
a pressure floor, which ensures that the Jeans scale is resolved by at least 
7 cells \citep{CDB}.  
Star formation is assumed to occur at densities above a threshold of $1\cmc$ 
and at temperatures below $10^4$K. 
More than 90\% of the stars form at 
temperatures well below $10^3$K, and more than half the stars form at 300~K.
The code implements a stochastic star-formation model that yields
 the empirical Kennicutt-Schmidt law \citep{Kennicutt98}. 

The base model incorporates thermal stellar feedback, in which the combined 
energy from stellar winds and supernova explosions is released as a constant 
heating rate over $40\Myr$ following star formation, the typical age of the 
lightest star that explodes as a core-collapsed supernova. 
The heating rate due to feedback may or may not overcome the cooling 
rate, depending on the gas conditions in the star-forming regions (CK09). No shutdown of cooling is implemented. 
We also include the effect of runaway stars by assigning a velocity kick of 
$\sim 10 \kms$ to 30\% of the newly formed stellar particles. 
The code also includes the later effects of Type-Ia supernova and 
stellar mass loss, and it follows the metal enrichment of the ISM. 


This base model has a few minor modifications with respect to the original CK09 model, used in \citet{CDB,Ceverino12}.
First, the current version uses a Chabrier IMF \citep{Chabrier05}, as opposed to the original Miller-Scalo IMF.
The star formation efficiency is lower by a factor of 3 with respect to the original CK09 value, which was used in simulations of lower resolution \citep{CDB,Ceverino12}.
At the resolution of these new simulations, this value roughly yields the empirical Kennicutt-Schmidt law  \citep{Kennicutt98}.
See Appendix A for more details.
Throughout the rest of the paper, we will refer to this base subgrid model as the run without radiation pressure (``NoRadPre" run).

\subsection{Modeling of Radiation Pressure}
\label{sec:RadPre}

The modeling of radiation pressure is done through the addition of a non-thermal pressure term,
 $P_{\rm rad}$, to the total gas pressure in regions where ionizing photons from massive stars are produced and trapped.
This ionizing radiation injects momentum around massive stars, pressurizing star-forming regions, as described in Appendix B of \cite{Agertz12}.
This extra pressure should be isotropic, such that the sum of thermal and radiative pressures also remains isotropic.
We assume an isotropic radiation field within a given cell, so the intensity, $I$, is constant and the radiation pressure is given by
\begin{equation}
P_{\rm rad} = \frac{4 \pi} {3 c} I. 
\end{equation}
This is equivalent to assuming that the radiation pressure is one third of the radiation energy density.
Finally we need a model for the intensity of radiation. 
If we assume a distribution of sources with a luminosity $L=\Gamma'  m_*$, where $m_*$ is the mass of stars and 
$\Gamma'$ is the luminosity of ionizing photons per unit stellar mass, 
the intensity is
\begin{equation}
I =  \frac{ \Gamma'  m_*} { A},
\label{eq:I}
\end{equation}
where $A$ is the area of the region enclosing the sources of radiation. It could be a sphere that contains the optically-thin region around the star cluster. This uncertainty in the geometry translates to a factor of a few in the accuracy of  the calculation.
If we assume that radiation is coming from a radiating sphere of radius $R$, $A=4 \pi R^2$, the radiation pressure can be expressed as
\begin{equation}
P_{\rm rad}= \frac{ \Gamma'  m_*}{R^2 c}, 
\label{eq:RadPre}
\end{equation}
where we absorb a factor of $\sqrt 3$ in the value of the radius, $R$, which is set to half the cell size for a cell hosting a stellar  mass $m_*$ and it is equal to the cell size for the closest neighbors of that cell.
This final expression is accurate to within a factor of a few.
Variations of up to a factor of 4 in $\Gamma'$ do not affect our results (Appendix B).

The value of $\Gamma'$ is taken from the stellar population synthesis code, \textsc{starburst99} \citep{Leitherer99}. We use a value of $\Gamma'= 10^{36}$ erg s$^{-1} \msun^{-1}$, which corresponds to the time-averaged luminosity per unit mass of the ionizing radiation during the first 5~Myr  of evolution of a single stellar population. After 5~Myr, the number of ionizing photons declines significantly, as the main sources, 
very massive stars
 explode as supernovae.  
The importance of radiation pressure also depends on the optical depth of the gas within a cell.
The gas should be optically thick to ionizing radiation.
Otherwise, that radiation can escape freely, 
without exerting any significant pressure on the gas. 
We use a hydrogen column density threshold, $N_{\rm rad}=10^{21} \cms$, above which ionizing radiation is effectively trapped and radiation pressure is added to the total gas pressure. This value corresponds to the typical column density of cold neutral clouds, which host optically-thick column densities of neutral hydrogen,  $N_{\rm HI}>10^{20} \cms$, assuming a UV opacity of $\kappa_{\rm UV}\sim10^3 \cms {\rm g}^{-1}$ \citep{ThompsonQuataertMurray05}.  

In summary, our current implementation of radiation pressure adds the value of radiation pressure given in \equ{RadPre} to the total gas pressure in the cells (and their closest neighbors) that contain stellar particles younger than 5~Myr and whose column density exceeds $10^{21} \cms$.
If these two conditions are not fulfilled, there is no extra pressure coming from radiation.

\subsection{Including Radiation Effects on Gas Cooling and Heating}
\label{sec:LS}

\begin{figure} 
\includegraphics[width =0.49 \textwidth]{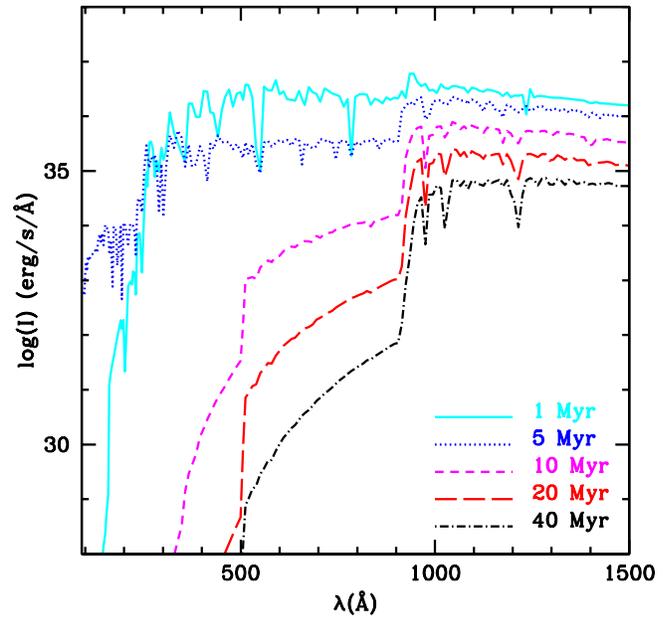}
\caption{Stellar spectra of a single stellar population with $\Ms=10^3 \ \msun$, solar composition and different ages, computed using \textsc{starburst99}.
The spectrum gets significantly harder at 5~Myr due to the contribution of WR stars.
The ionizing radiation, below 912 \Ams \ is significantly suppressed after 10 Myr.}
\label{fig:spectrum}
\end{figure}

\begin{figure*} 
\includegraphics[width =0.99 \textwidth]{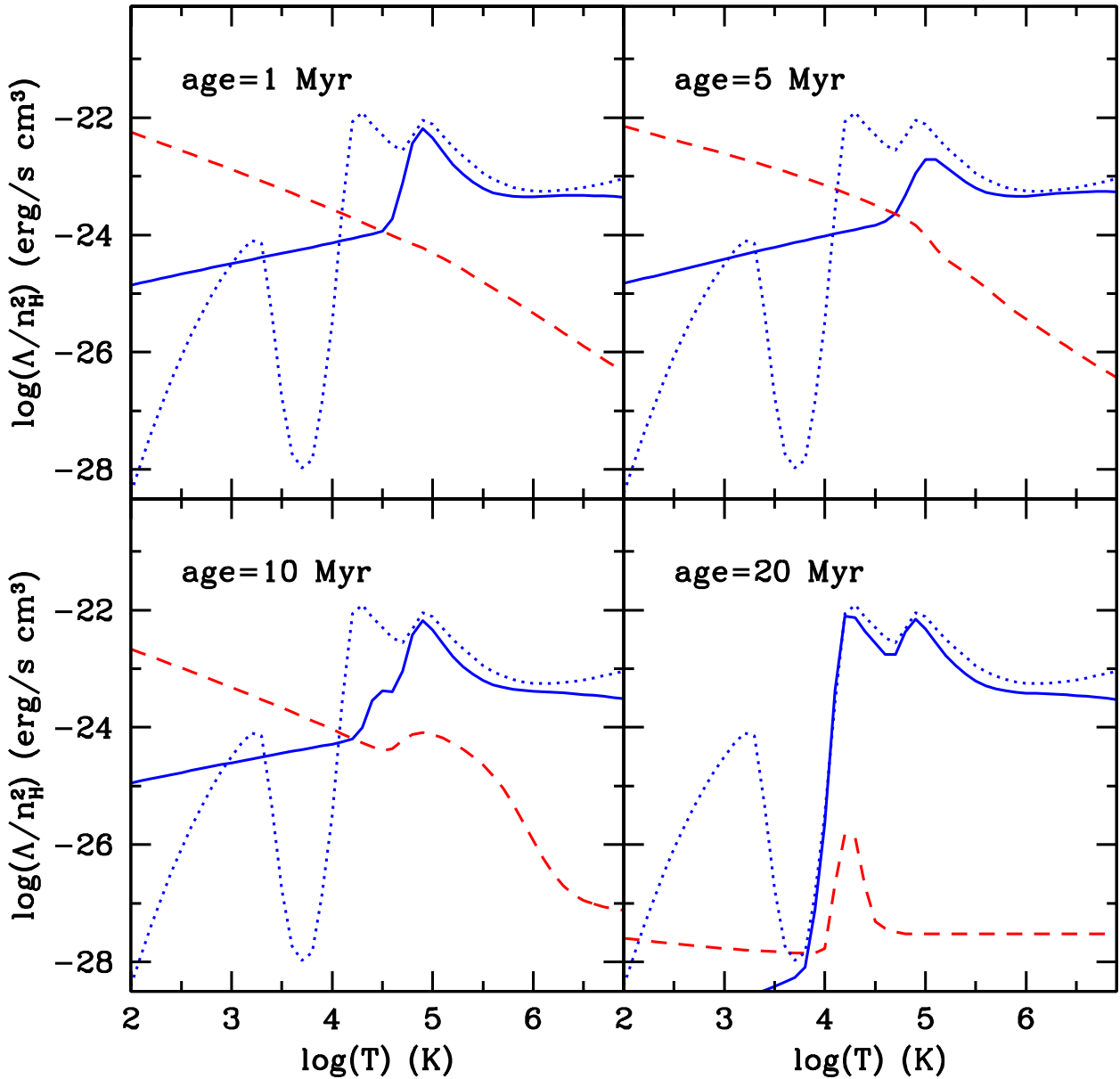}
\caption{\textsc{cloudy} cooling (blue solid curve) and heating (red dashed curve) rates for gas density $n=10 \cmc$ and  gas metallicity $Z=10^{-3} Z_{\odot}$, illuminated by a star cluster of $\Ms=10^6 \ \msun$.  The four panels correspond to different ages of the stellar population. For comparison, the dotted blue curves correspond to the cooling of UV shielded gas. Photoionization drastically reduces the cooling around T$=10^4-10^{4.5}$ K, if the number of ionizing photons is enough to 
ionize almost all hydrogen atoms.
Photoheating significantly increases the temperature and thermal pressure of the HII region surrounding the star cluster by a factor of about 100.}
\label{fig:CHage}
\end{figure*}

\begin{figure*} 
\includegraphics[width =0.99 \textwidth]{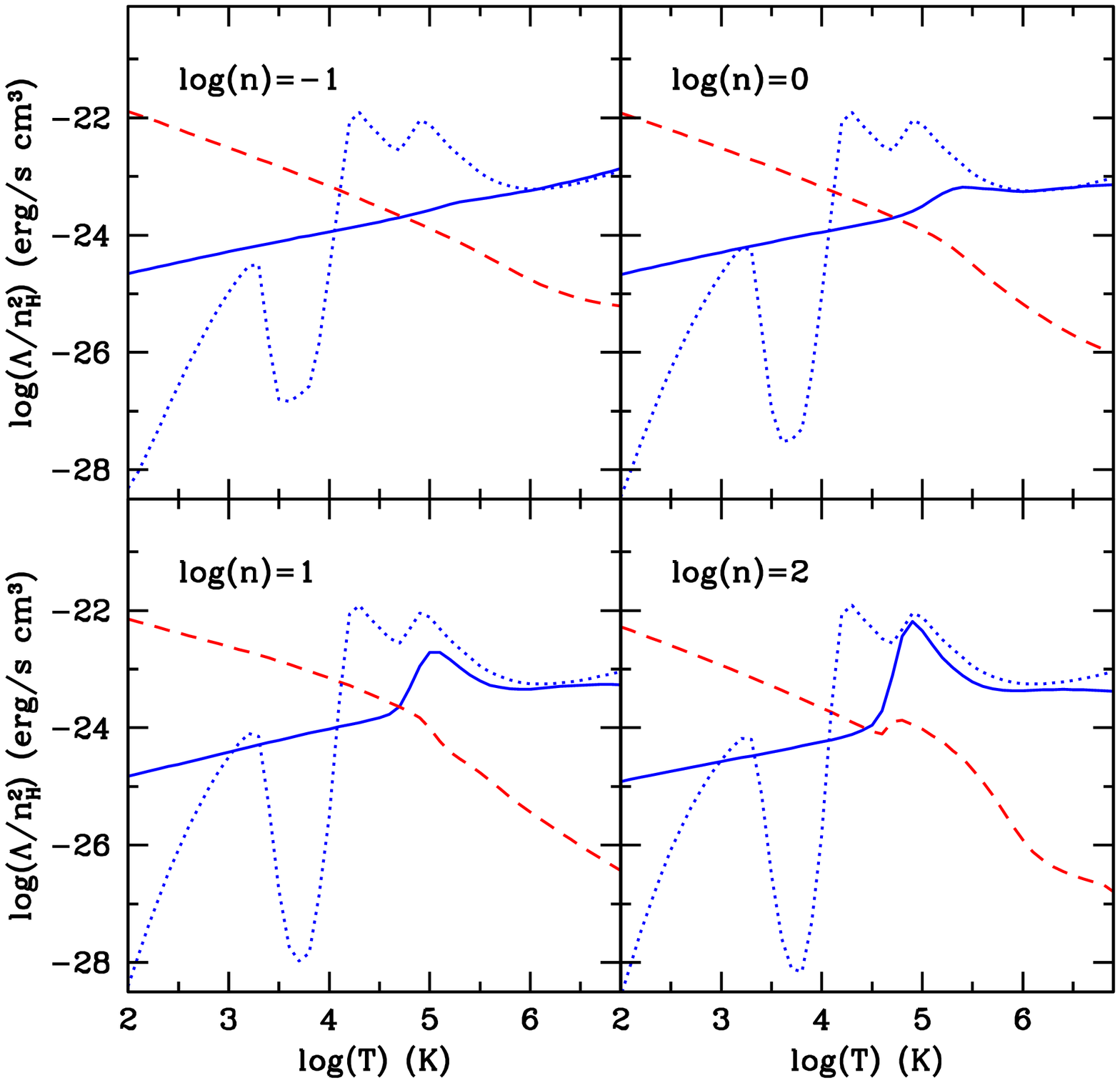}
\caption{Same as Figure 2, with a stellar age of 5 Myr, where each panel corresponds to a different gas density.
 Higher densities correspond to higher column densities, which are able to attenuate the incident radiation. This changes the ionization conditions and the corresponding cooling curves.}
\label{fig:CHden}
\end{figure*}

In addition to radiation pressure, the local UV flux from young star clusters also affects the cooling and heating processes in star-forming regions through photoheating and photoionization.
These processes depend on the mass and age of the star cluster.
The ionizing UV flux of a single stellar population (SSP) depends on its age because ionizing photons are mainly produced in
massive main-sequence stars and post-main-sequence stars, such as Wolf-Rayet (WR) stars.
They have lifetimes of 1-5~Myr.
Therefore, the UV stellar spectrum of a SSP varies significantly during the first few Myr.

We model this age variation using the \textsc{starburst99} code.
Five different spectral shapes of SSP of different ages are considered: 1, 5, 10, 20, 40 Myr.
\Fig{spectrum} shows these spectra for a SSP of 10$^3 \ \Msun$.
The main difference between the spectra of different ages occurs in the regime of ionizing radiation, below  912 \Ams.
The 1-Myr-old spectrum shows significant ionizing radiation in the EUV regime, up to $\sim 200$ \Ams.
After 5 Myr, the overall luminosity has decreased, but the radiation is significantly harder, extending into the soft X-ray regime, up to $\sim 100$ \Ams. This is due to the presence of WR stars.
By 10 Myr, the level of ionizing radiation has been severely reduced by more than 2 orders of magnitude with respect to the 1-Myr-old spectrum. This reduction continues,
as massive stars explode in core-collapsed supernovae.
The stellar ionizing flux also depends on the metallicity of the stars, but this is a secondary effect that we are ignoring for the moment.

These stellar spectra are used as incident radiation on the \textsc{cloudy}  models (v. 8),
which generates tables for the cooling and heating rates as a function of the gas density, temperature, and metallicity, as well as incident spectra. 
We consider the 5 spectral shapes discussed above, 
as functions of stellar age and 4 different normalizations, corresponding to SSP of 4 different masses: $10^3$, $10^4$, $10^5$, $10^6 \ \msun$.
This yields 20 different spectra, as functions of stellar mass and age.
 We adopt a 25 pc thick slab, the typical resolution of these simulations. The slab is illuminated from one side and the cooling and heating rates are computed at the back side of the slab.

\Fig{CHage} shows an example of the cooling and heating rates for different stellar ages. This example assumes a gas of relatively high density, $n=10 \cmc$, and almost primordial composition,  $Z=10^{-3} Z_{\odot}$. This gas is illuminated by a super star cluster of $\Ms=10^6 \ \msun$, a typical massive cluster in starburst galaxies \citep{Fall04, Turner00}.
The dotted curves in \Fig{CHage} correspond to the cooling rates for the UV-shielded gas. 
As the composition is close to primordial, only the peaks that correspond to H$_2$ and to the collisional excitation of neutral H and He are visible.
By contrast, if the same material is close enough (25 pc) to a 1-Myr-old massive star cluster, the gas cooling is dramatically 
 affected: the HI peak of the cooling curve has disappeared. 
The strong stellar radiation is able to ionize almost all hydrogen atoms, effectively suppressing HI cooling.
However, the He peak of the cooling curve remains unmodified, because the ionization of helium needs a much harder spectrum.
This is partially achieved after 5 Myr, when the harder spectrum from WR stars is able to significantly reduce He cooling.
With increasing stellar age, the cooling curves slowly revert back to the UV-shielded case, as the number of ionizing photons declines.

The effects of photoionization are complemented by photoheating, which sets 
the heating rate equal to the cooling rate at
an equilibrium temperature of \Teq$= 10^4$ -10$^{4.5}$ K. 
The net effect is an increase in the temperature and thermal pressure by a factor of about 100
with respect to the temperature of dense, UV-shielded gas ($T\simeq300$ K).
Due to the fact that the cooling times are very short in these dense regions, the gas settles into an equilibrium between cooling and heating on time scales much shorter than the local dynamical time.

Cooling and heating rates depend not only on the properties of the incident ionizing radiation, but also on the properties of the gas.
In particular, the gas density controls the attenuation of the incident UV flux.
If the gas column density is high enough, the gas can shield itself against the local stellar radiation.
This is illustrated in \Fig{CHden}, which shows the cooling and heating rates of gas at different densities, 
but with the same radiation coming from a 5-Myr-old cluster of $10^6 \ \msun$.
Even in the case of high densities, $n=100 \cmc$, the HI peak of the cooling curve has disappeared. 
This effect can be very important in the overheating and pressurization of actively star-forming and dense regions.

%
In more abundant star clusters of lower masses, such as Orion-like clusters ($\Ms \sim 10^3 \ \msun$), the effect of radiation on the cooling and heating rates is less dramatic but still relevant.
A small cluster of $10^3 \ \msun$ is able to heat the surrounding dense gas ($n=10 \cmc$) up to \Teq$=2 \times 10^4$ K, during 1 Myr.
The same equilibrium temperature can be maintained by a cluster of  $10^4 \ \msun$ during 5 Myr, the typical lifetime of observed HII  regions \citep{Lopez13}.
More massive clusters are able to heat gas with higher densities.
Finally, different gas metallicities give similar results, because metals with H-like 
 or He-like ionization potentials are as ionized as primordial gas,
when exposed to the same radiation field.

These new cooling and heating rates are implemented in the following way.
Each stellar particle of a given stellar mass and age enters into one of four bins in stellar mass: $\Ms<5 \times 10^3 \ \Msun$,  
$5 \times 10^3 <\Ms<5 \times 10^4 \ \Msun$, $5 \times 10^4 <\Ms<5 \times 10^5 \ \Msun$, $\Ms> 5 \times 10^5 \ \Msun$,  and in one of five bins in stellar age: ${\rm age}<3 \Myr$, $ 3 \Myr <{\rm age}<7.5 \Myr$, $ 7.5 \Myr <{\rm age}<15 \Myr$, $ 15 \Myr <{\rm age}<30 \Myr$, $ 30 \Myr <{\rm age}<40 \Myr$.
The cooling and heating rates of the cell containing that stellar particle are taken to be that corresponding to the mass and age bin to which the stellar particle belongs.
In the case that several stellar particles of different properties are found in the same cell, only the 
rates providing the highest \Teq \ are adopted.
This is motivated by the fact that the SSP 
yielding the highest equilibrium temperature usually generates the radiation 
that dominates the ionizing part of the spectrum.

\subsection{Including infrared Radiation}
\label{sec:IR}

The last addition to the subgrid model takes into account the boosting effect of the trapped infrared photons on radiation pressure.
It is a simple model in which the multiplicative effect of the infrared radiation is proportional to the gas density, or equivalently, to the gas surface density for a constant cell size, as in \cite{Hopkins11}.

If the gas density exceeds a threshold of 300 $\cmc$, the expression for radiation pressure is
\begin{equation}
P_{\rm IRrad}= (1+ \tau_{\rm IR} ) P_{\rm rad},
\label{eq:RadPre_IR}
\end{equation}
where $P_{\rm rad}$ is the radiation pressure of \equ{RadPre}, 
and $\tau_{\rm IR}$ is the optical depth for infrared radiation given by
\begin{equation}
\tau_{\rm IR}= \frac{n}{300 \cmc}.
\end{equation}
This simple model gauges the importance of infrared radiation in the efficiency of  star formation. Its results can be compared to other similar approaches \citep{Hopkins11, Agertz12,Aumer13}.

\subsection{Summary of runs}

Table \ref{tab:1} summarizes the different combination of models used in this paper.
The first run, NoRadPre, only includes the base model (\se{base}).
The RadPre run includes the model of radiation pressure (\se{RadPre}).
The RadPre\_LS run adds only the model of local sources of photoionization and photoheating (\se{LS}), while
the RadPre\_IR run adds only the boost of IR radiation (\se{IR}) to the model of radiation pressure.
Finally, the RadPre\_LS\_IR run includes both the effects of local heating and IR boosting.

\begin{table} 
\caption{Summary of runs}
\begin{center} 
 \begin{tabular}{lccc} \hline 
\multicolumn{2}{c} {Run } \ \  Radiation & Local sources of photoionization & IR  \\ & pressure & and photoheating & radiation\\
\hline 
NoRadPre &  NO & NO  & NO  \\ 
RadPre &  YES & NO  & NO  \\ 
RadPre\_LS & YES  & YES  & NO  \\ 
RadPre\_IR & YES & NO  & YES  \\ 
RadPre\_LS\_IR  & YES & YES  & YES  \\ 
\hline 
 \end{tabular} 
\end{center} 
\label{tab:1} 
 \end{table}

\section{Initial Conditions and Simulation Details}
\label{sec:IC}

The radiative feedback model, described in the previous section, has been tested in a cosmological simulation of the formation of a halo selected to have a maximum circular velocity of $V_{\rm c} \simeq 120 \kms$ at $z=0.8$.
The other selection criterion is that the halo shows no ongoing major merger
at $z=0.8$ This eliminates less than $10\%$ of the halos with similar maximum circular velocity,
and it has no noticeable selection effect at $z \sim 3$, where our main analysis
is performed.
The selected halo has a virial mass of $\Mv=4 \times 10^{11} \ \Msun$ at $z=0.8$
(the halo has a virial mass slightly lower than $10^{12} \ \msun$ today, slightly less massive than the Milky Way).  
At the analyzed snapshot at $z=3$, the virial mass is around $\Mv\simeq 10^{11} \ \msun$, a significant fraction of the mass of the remnant at $z=0.8$.

The dark-matter halo was drawn from a N-body simulation assuming a
$\Lambda$CDM cosmology with the WMAP5 parameters,
$\omm=0.27$, $\oml=0.73$, $\omb= 0.045$, $h=0.7$ and $\sigma_8=0.82$.
The selected halo
was filled with gas and refined to a much higher resolution on an adaptive
mesh within a zoom-in lagrangian volume that encompasses the mass within
twice the virial radius at $z =0.8$, roughly a sphere of comoving radius 0.5 Mpc \citep{Klypin02}.
The total comoving cosmological box is  40 Mpc/h. 
The galaxy has been evolved from an initial redshift, $z=100$, 
on an adaptive comoving mesh refined in the dense regions
to cells of minimum size between 17-35 pc in physical units.
The number of dark matter particles is $2.2 \times 10^7$, with a DM particle mass of $8.3\times 10^4 \ \msun$.
The particles representing star clusters have a minimum mass of $10^3\  \msun$, similar to the stellar mass of an Orion-like star cluster.
The dark matter mass resolution of this simulation is slightly better than the Eris simulation \citep{Guedes11, Shen13} and slightly worse than level 3 resolution of the Aquarius suite of N-body-only runs \citep{Springel08}.

Each AMR cell is split into 8 cells once it contains a mass in stars and 
dark matter higher than $2.6\times 10^5 \ \msun$, equivalent to three dark-matter 
particles, or once it contains a gas mass higher than $1.5\times 10^6 \ \msun$. This 
quasi-lagrangian strategy ends at the highest level of refinement 
that marks the minimum cell size at each redshift. 
The maximum spatial resolution is 108 pc in co-moving units.
In particular, the minimum cell size becomes 27 pc in physical units by $z=3$, when most of the analysis described in this paper has been performed.

\section{Results at redshift $z$=3}
\label{sec:results}

 \begin{figure} 
\includegraphics[width =0.47 \textwidth]{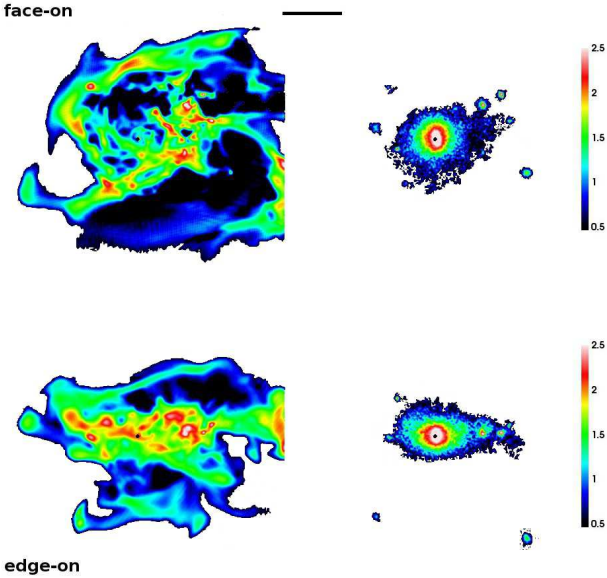}
\caption{Projected mass density maps of the NoRadPre run at z=3. Gas (left) and stars (right), viewed face-on (top) and edge-on (bottom).   Each panel is 20 kpc. The horizontal bar represents 4 kpc and the small dot marks the galaxy centre. The color scale represents the surface density in units of $\log ( \Msun \pc^{-2})$.}
\label{fig:SFNEW}
\end{figure}

\begin{figure} 
\includegraphics[width =0.49 \textwidth]{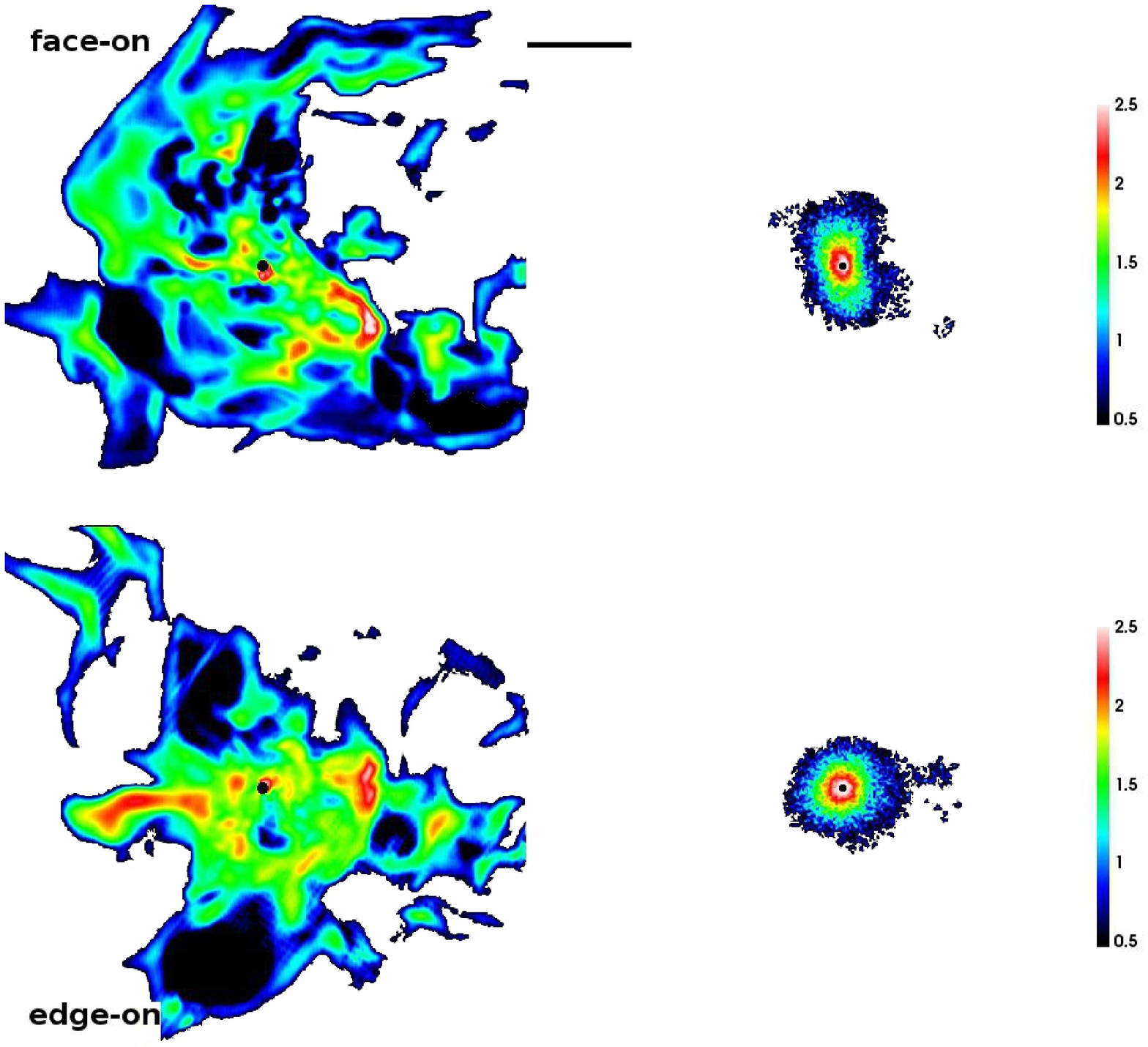}
\caption{Same as \Fig{SFNEW} for the RadPre\_LS run.}
\label{fig:LSNEW}
\end{figure}

The analysis has been performed at redshift $z=3$, 
when the main progenitor has a virial mass of $\Mv \simeq 1.3 \times10^{11} \ \Msun$ and a virial radius of $\Rv\simeq 42 \kpc$.
These virial properties are roughly the same in all runs. There is a maximum deviation of 14\% in the value of the virial mass and only 4\% for the virial radius in the different runs at $z=3$. Therefore, the inclusion of radiative feedback does not significantly alter the virial properties of the halo, which are governed primarily by the growth of large-scale density perturbations and dark-matter accretion into halos. These are processes on which feedback has little effect.

\subsection{ The Effect of Radiation Pressure on the Gas and Stellar Distributions}

\Fig{SFNEW} and \Fig{LSNEW} compare the NoRadPre run with the RadPre\_LS run, but all runs with radiation pressure show similar features in their distributions of gas and stars.
In all runs, even without radiation pressure, the gas is distributed in an irregular thick disc.
The combination of high resolution (27 proper pc) and feedback produces clumpy, multiphase gas in galaxies of this mass scale (see Figure 11 in CK09). 
The gas within the galaxy is distributed in dense elongated features with clouds embedded in more diffuse medium. 

When radiation pressure is taken into account, along with thermal feedback, the gas distribution changes substantially. The disc looks even more perturbed, probably due to an increase in the turbulence driven by feedback.
The diffuse gas with relatively low column densities, $\Sigma_{\rm Gas}\sim10^{1.5} \ \Msun \pc^{-2}$, is more extended.
The most compact clouds with high column densities, $\Sigma_{\rm Gas} \geq 100 \ \Msun \pc^{-2}$, have mostly disappeared.
 They are dispersed quickly by the radiation pressure exerted on the gas in star-forming regions.
 
 It is important to stress that these compact clouds are not the typical giant clumps found in more massive, gravitationally unstable discs at high redshift \citep{Ceverino12}. The clouds in this low-mass disc have much
 lower masses ($10^7 \ \Msun$) than the kpc-scale giant clumps ($10^8-10^9 \ \Msun$). 
It is much more difficult to unbind giant clumps by star-formation-related processes \citep{KrumholzDekel}.
In contrast, the clouds formed in this simulation are similar to local GMCs, which are quickly dispersed in few crossing times.

If the average gas surface density is much lower in the run with radiation pressure, the average star formation rate should also be lower, because all runs roughly obey the Kennicutt-Schmidt relation.
Therefore, the stellar surface density should also decrease. Indeed, the
stellar distribution in the run with radiation pressure is less concentrated and reaches lower central densities.
The stellar half mass radius increases from 0.7 kpc (NoRadPre) to 1.5 kpc with the inclusion of radiation pressure.
All small stellar clumps that appear in the NoRadPre run have disappeared when radiation pressure is implemented. In particular, the small satellite galaxies off the disc plane have been quenched because star formation is very inefficient in these small galaxies, due to radiation pressure and photoheating.

The stellar distribution in the  RadPre\_LS run resembles a prolate structure.
In the face-on view, based on the galaxy angular momentum, the minor-to-major axial ratio is about $b/a=0.5$, measured at the isodensity contour of $10^{1.5} \ \Msun \pc^{-2}$. 
This irregular morphology is also observed in some low-mass, high-z galaxies of $\Ms\simeq10^9 \ \Msun$ \citep{vanderWel12}, although the observed distribution of galaxy shapes is quite broad.
Better simulation statistics are needed for a fair comparison of distribution of galaxy shapes in high-z samples.
In this particular case,
the morphology is far away from a round, circularly symmetric disc. 
This is a consequence of the low efficiency of star formation, driven by radiation pressure.
Star formation does not occur across the gaseous and turbulent disc, yielding a stellar disk.
Instead, star formation proceeds only in dense clouds along a bar-like structure.

\subsection{Low Efficiency of Star Formation}

\begin{table*} 
\caption{Galaxy properties at $z=3$: galaxy stellar mass ($\Ms$), stellar-to-virial mass ratio (f$_{\rm stars}$), halo baryonic fraction (f$_{\rm Bar}$), star formation rate (SFR) in $\Msun$ yr$^{-1}$, specific star formation rate (SSFR) in Gyr$^{-1}$, gas mass within the virial radius (M$_{\rm Gas,Halo}$), gas mass within the galaxy (M$_{\rm Gas}$), and star-forming gas within the galaxy (M$_{\rm SF-Gas}$). All masses have units of $\Msun$.}
 \begin{center}       
 \begin{tabular}{lcccccccc} \hline     
\multicolumn{2}{c} {Run} \ \ \ \ \ \ \ \ \ \ \ \  $\Ms$  \ \ \ \ \ \ \ \  &  f$_{\rm stars}$   & f$_{\rm Bar}$  & SFR & SSFR & M$_{\rm Gas,Halo}$ &  M$_{\rm Gas}$  & M$_{\rm SF-Gas}$  \\\hline 
NoRadPre &  $2.50    \times 10^9$ &                                           0.0168      &             0.12 &             3.22 &  1.29 &  $1.58 \times 10^{10}$   &   $ 1.2 \times 10^9$       & $5.7 \times 10^8$   \\
RadPre          & $1.02  \times 10^9$ &                                            0.0075      &             0.13 &               0.82 &  0.81 &  $1.60 \times 10^{10}$   &   $ 0.82 \times 10^9$     &  $2.7 \times 10^8$   \\                  
RadPre\_LS  & $0.93 \times 10^9$ &                                            0.0067      &             0.13 &                1.40 &  1.51 &  $1.74 \times 10^{10}$   &   $ 0.93 \times 10^9$     &  $2.3 \times 10^8$  \\
RadPre\_IR   & $0.83 \times 10^9$ &                                            0.0062      &             0.11 &                0.95 &  1.14 &  $1.45 \times 10^{10}$   &   $ 0.75 \times 10^9$     & $2.1  \times 10^8$  \\
RadPre\_LS\_IR & $0.86 \times 10^9$ &                                      0.0064      &             0.12 &                1.22 &  1.41 &  $1.47 \times 10^{10}$   &   $ 0.96 \times 10^9$     & $3.8 \times 10^8$         
 \end{tabular} 
 \end{center}
\label{tab:3}
 \end{table*}

\begin{figure} 
\includegraphics[width =0.49 \textwidth]{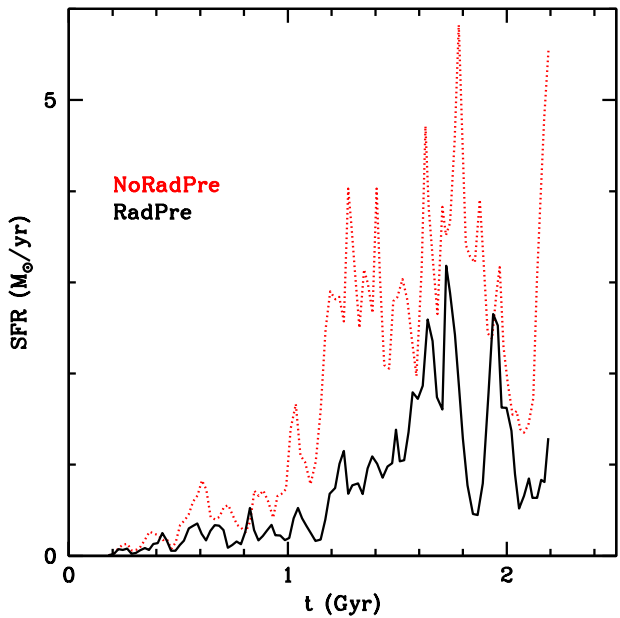}
\caption{Star formation rate history with radiation pressure (solid curve) and without radiation pressure (red dotted curve).
Radiation pressure significantly reduces the star formation in the small galaxy progenitors at high redshift.}
\label{fig:SFR}
\end{figure}

In the previous section, we claimed that radiative feedback significantly affects star formation by reducing the size of regions with high surface densities and high star formation rates.
\Fig{SFR} compares the SFR history of  the NoRadPre and RadPre runs.  The SFR history is measured using the formation time of all stars within a radius of 5 kpc from the galaxy centre at $z=3$.
Therefore, the SFR history is a compilation of the SFR in the main progenitor at $z\simeq3$, as well as in all small previous progenitors that also formed stars at higher redshifts.

In general, the SFR in the case with radiation pressure is a factor of $\sim$3 lower than in the NoRadPre run.
Even the high-SFR peaks, related to starburst events, have much lower SFR values 
when radiation pressure is taken into account.
This is consistent with the findings from isolated disc simulations \citep{Hopkins11,Agertz12}.
For example, the isolated, MW-like disc in \cite{Agertz12} has a factor 2 difference in the SFR history between the runs with and without radiation pressure.
However, these non-cosmological simulations can not address the cumulative effect of this radiation-regulated star formation on the formation of galaxies.
The other runs that include photoheating and/or infrared trapping, show SFR histories similar to that in the RadPre case.

\tab{3} lists the properties of the galaxy residing in the selected halo at $z=3$ in the different runs. 
All galaxy properties are computed within 10\% of the virial radius, which is roughly equal to 4 kpc.
The addition of radiation pressure reduces the stellar mass of the central galaxy by a factor of 2.5. Other effects, such as IR trapping (RadPre\_IR) or photoheating from local sources (RadPre\_LS) decrease the stellar mass slightly further up to a factor 3 lower than in the case without radiation pressure. 
This means that radiation pressure from ionizing photons regulates galaxy growth, although other processes also contribute to keep a low efficiency of star formation in $\Mv\simeq10^{11} \ \Msun$ halos at high redshifts.

One measure of this low efficiency of galaxy formation is the stellar-to-virial mass ratio, f$_{\rm stars}$=0.6\% in the RadPre\_IR run. 
This is consistent with current abundance matching models \citep{Moster12, Behroozi12} within the systematic uncertainties.
However, the values from the simulations are on the high side of the allowed range, f$_{\rm stars}\simeq 0.06\%-0.6\%$ for halos with masses of $\Mv\simeq10^{11} \ \Msun$ at $z=3$.
This scatter is large, so better abundance matching models with better observational constraints at $z\simeq3$ will be needed to clarify the issue of low galaxy efficiency at these high redshifts.
In contrast, other simulations without radiative feedback report much higher stellar fractions. For example, the Eris2 simulation \citep{Shen13} has 10 times higher stellar mass, f$_{\rm stars}$=6\%, for a  similar halo mass, $\Mv=2.6 \times 10^{11} \ \Msun$ at a similar redshift, $z=2.8$.
Only simulations that include radiative feedback provide low stellar fractions.
The closest results are shown in \cite{Aumer13}, who report f$_{\rm stars}$=0.6-0.8\% for $\Mv\sim 10^{11} \ \Msun$ at $z=3.5$.
This is consistent with our findings, although their model of radiative feedback is quite different.

The star formation rate also decreases substantially, by a factor $\sim$4, due to the effect of radiation pressure. 
In these simulations, a value of SFR$\simeq$1$\sy$ is the typical value for a galaxy living in a $10^{11} \Msun$ halo at $z=3$.
This is roughly consistent with halo abundance matching models \citep{Behroozi12}. 
It is much lower than the $ 20  \sy$ found by \cite{Shen13}, and it is slightly lower than the $2-3 \sy$ reported in \cite{Aumer13} for a similar halo mass and redshift. 
Finally, the specific star formation rate in the RadPre run, SSFR=SFR/$\Ms$, is lower by a factor of 1.6. 
This decrease is less dramatic than the decrease in SFR
because both the stellar mass and the SFR are lower with the addition of radiative feedback.

The decrease in the star formation rate is correlated with a decrease in the amount of cold ($T<10^4$ K) and dense ($n>1 \cmc$) gas available for star formation within the galaxy.
Without radiation pressure, this
star-forming gas accounts for almost 50\% of all the gas in the galaxy.
This is a feature of the overcooling regime, in which 
a large fraction of the galactic gas is able to form stars.
When radiation pressure is considered, the amount of star-forming gas 
decreases to 30\% of all galactic gas.
This fraction is reduced to 25\% when photoionization and photoheating are included.
Therefore, radiative feedback regulates the star formation process by regulating the amount of cold and dense star-forming gas.

\subsection{Cumulative Gas Profiles}

\begin{figure*} 
\includegraphics[width =0.9 \textwidth]{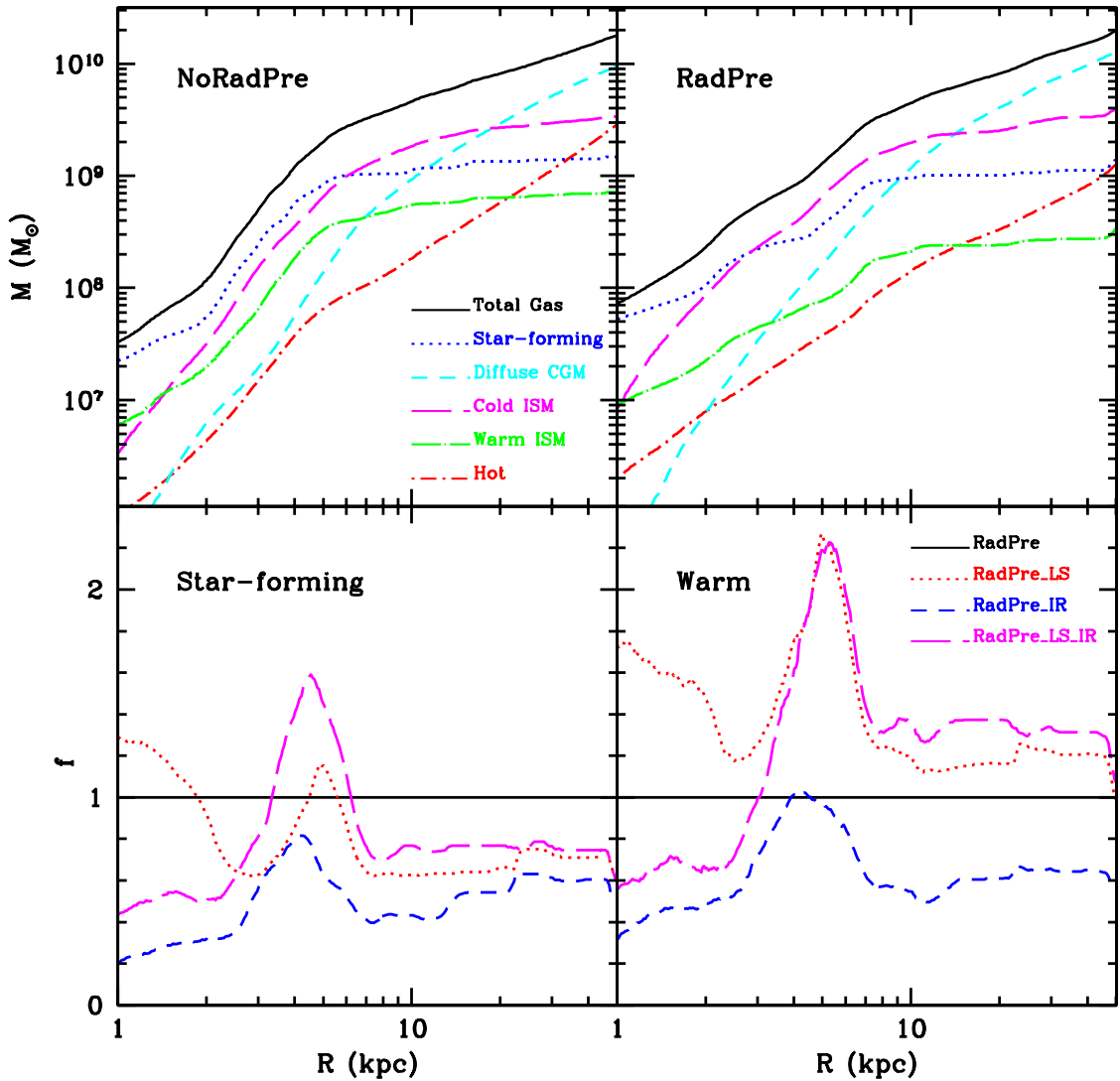}
\caption{
Top panels: cumulative mass profiles of the different gas components (see main text) for NoRadPre (left) and RadPre (right).
Radiation pressure decreases the amount of cold and dense star-forming gas within the galaxy, but
the diffuse gas, which dominates the gas mass on halo scales, is mainly unaffected by radiative feedback.
Bottom panels: Cumulative mass of star-forming gas (left) and warm ISM (right) in the different runs, normalized to the values in RadPre. The infrared boost to radiation pressure decreases the amount of star-forming gas and photoheating increases the amount of warm and relatively dense gas.}
\label{fig:GasProf}
\end{figure*}

The distinct gas phases are affected by radiative feedback in different ways.
\Fig{GasProf} shows the cumulative mass profiles of the following gas components: hot gas ($T>10^5$ K), star-forming gas ($n>1 \cmc$ and $T<10^4$ K), diffuse and warm circumgalactic medium (CGM) ($n<0.1 \cmc$ and $T<10^5$ K), slightly denser and colder gas ($T<10^4$ K and $0.1 \cmc < n < 1 \cmc$), and warm interstellar gas ($10^4 \ \rm{K} < T < 10^5 \ \rm{K}$  and $n > 0.1 \cmc$). 
In all runs, the star-forming gas dominates the gas reservoir inside a given radius, $R_{\rm SF}$.
However, the value of this radius depends on radiative feedback.
Without radiation pressure, star-forming gas dominates the gas reservoir throughout the galaxy, $R_{\rm SF}=6$ kpc.
With the inclusion of radiation pressure, that radius is reduced to $R_{\rm SF}=2-3$ kpc.
Similar values are found in other runs with radiation pressure.
In these runs, the main gas component within the galaxy ($R<4-5$ kpc) is the relatively dense and cold, non-star-forming gas.

Intermediate density gas also dominates the gas budget in the inner halo, up to $\sim15$ kpc ($\sim36$\% of the virial radius).
More diffuse gas becomes the main gas component outside this radius.
It is important to notice the large reservoir of gas at relatively large radii ($R>4 \kpc$).
The halo retains a lot of gas at these redshifts, M$_{\rm Gas,Halo}(r<\Rv)\simeq10^{10} \ \Msun$ (\tab{3}).
Most of the gas within the virial radius, mostly diffuse circumgalactic gas, is mainly unaffected by radiation pressure.
This means that the decrease in stellar mass, driven by radiation effects,
is not due to massive halo blow-outs. 
These large amounts of gas in the halo could in principle account for most of the observed high-z Damped Lyman-$\alpha$ systems in halos with a maximum circular velocity of $100 \kms$ \citep{ProchaskaWolfe97}. On the other hand,
 the amount of gas within the central galaxy ($R < 4$ kpc) is significantly lower when radiation pressure is taken into account, and the galaxy is gas rich, M$_{\rm Gas}/ \Ms \simeq 1$. 
Therefore, a significant amount of gas should be ejected from the galaxy, i.e., from the inner halo.
A more detail analysis of these outflows will be reported in a companion paper in preparation.

Massive outflows have been traditionally invoked to explain the low galaxy formation efficiency \citep{Dave11, Haas12}.
Outflows may eject a significant fraction of the ISM in low-mass halos. This limits the SFR and stellar mass.
Indeed, cosmological simulations of low resolution but large volumes has shown relatively low stellar-to-virial mass ratios, thanks to the implementation of different prescriptions of phenomenological wind models. \citep{SpringelHernquist03,
 OppenheimerDave06, OppenheimerDave08, OppenheimerDave10, Schaye10,Dave11,Haas12,Vogelsberger13}.
As an example, the stellar-to-virial mass ratios reported in \cite{Haas12} for their fiducial model are close to  f$_{\rm stars}$=1.2\% for halos with masses of $\Mv\simeq10^{11} \ \Msun$ at $z=2$.
\cite{Torrey13} report f$_{\rm stars}$=0.4-0.7\% at $z=2$ for the same halo mass. 
These values are similar to the stellar fractions reported in this paper.
However, a low stellar mass fraction can also be achieved if most of the gas stays in the halo but in the form of diffuse halo gas, which does not form stars. 
In the simulations, the baryonic mass fraction (gas+stars) within the halo is f$_{\rm Bar}\simeq12\%$, which is $\sim70$\% of the universal baryonic fraction. This is a large value, in comparison to the stellar mass fraction, f$_{\rm stars}\simeq0.7\%$, mentioned above.
Previous simulations with ad-hoc outflows produce a significantly lower fraction of baryons inside the galactic halo,  f$_{\rm Bar}\simeq6\%$ \citep{Haas12}.
This is a factor of 2 lower than the values found in our simulations for the same halo mass range.

The two variations of radiative feedback, infrared boosting and photoheating/photoionization, mostly affect the star-forming gas and the warm but relatively dense gas (\Fig{GasProf}).
With the infrared boost to radiation pressure, the amount of star-forming gas is reduced by $\sim$60\% in the inner galaxy ($R<3 \kpc$).
The star-forming gas accumulates at larger radii, $R\simeq4 \kpc$,  where it is only $\sim$20\% lower than in the RadPre run.
The effect of infrared boosting on the warm gas is nearly the same as that on the star-forming gas.
Photoheating mostly affects the warm gas. Its mass is up to a factor 2 higher than in the RadPre run at $R\simeq5 \kpc$.
The hot gas shows similar results, although its contribution to the total mass budget within the galaxy is significantly lower. This accumulation of warm/hot gas in the inner halo, is due to the extra heating coming from ionizing photons.

\subsection{Rising Circular Velocity Profiles}

\begin{figure} 
\includegraphics[width =0.49 \textwidth]{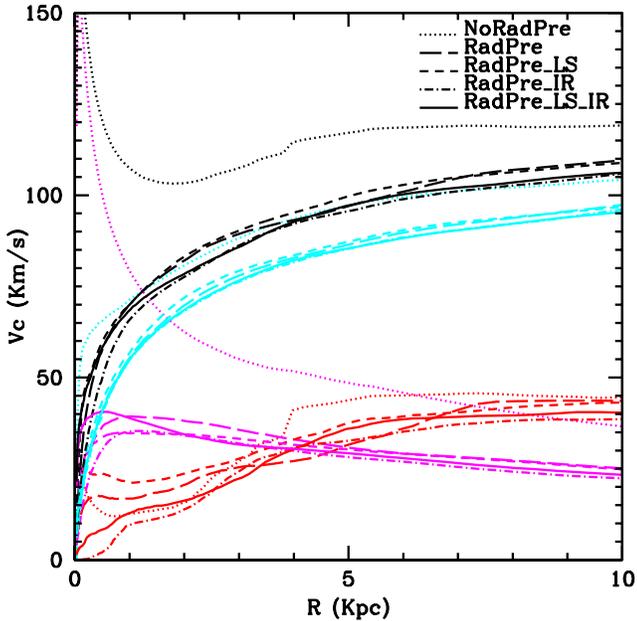}
\caption{Circular velocity profiles for all different runs. Each total profile (black) is decomposed in the stellar (magenta), gas (red) and dark matter (cyan) contributions.
Radiation pressure effectively reduces the overcooling problem and it prevents the formation of a dense and massive stellar core. 
The net result is a rising profile, dominated by the dark matter contribution.}
\label{fig:Vc}
\end{figure}

The distribution of stars within the galaxy depends on feedback.
Radiation pressure effectively pressurizes star-forming regions and promotes the expansion of overpressure bubbles that enhance the multiphase nature of the ISM. 
This prevents catastrophic overcooling, which generates 
runaway star formation and the formation of a small and dense galaxy, as discussed in CK09.
The profile of the circular velocity, $V_c=\sqrt{GM/R},$ where $G$ is the gravitational
constant and $M$ is the mass inside a radius $R$, illustrates this effect (\Fig{Vc}).

Without radiation pressure, the total circular velocity profile shows a strong peak at small radii. This is produced by a very concentrated (and massive) stellar distribution and is the typical profile in the overcooling regime (see top panel of Figure 12 in CK09).
If radiation pressure is taken into account, the stellar distribution is much more extended.
This results in a rising circular velocity profile, which is mostly dominated by the dark matter contribution, except at very small radii, $r\leq0.3 \kpc$, where stars dominate the mass.
This gentle rise is more consistent with the rotation curves of observed galaxies of similar mass \citep{PersicSalucci96}.

Similar results are found when other runs are compared.
The stellar distribution decreases in the RadPre\_IR and RadPre\_LS runs, but the maximum circular velocity of the stellar distribution differs only by 12\% from the RadPre run.
It seems that the different models of radiative feedback converge to a similar distribution of stars 
with differences of the order of 10\%-20\%
within the galaxy.
Similar results are also achieved in runs with different values for the parameters of the radiation pressure model (Appendix B).
Therefore, these results are robust against various modifications in the model of radiative feedback.

\subsection{Density-Temperature Diagrams}

\begin{figure*} 
\includegraphics[width =0.99 \textwidth]{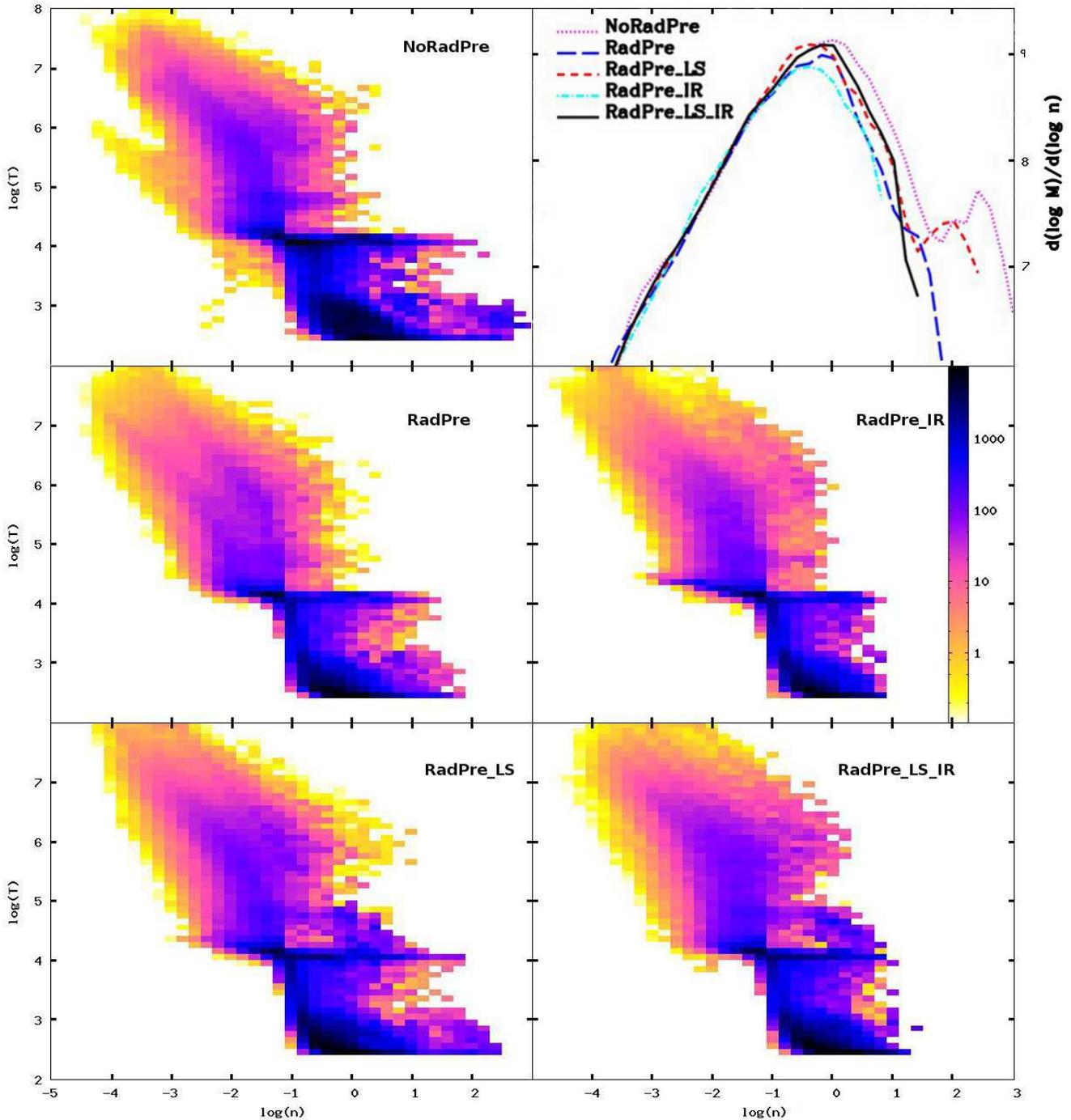}
\caption{Temperature versus density histograms of all cells within 5 kpc from the galaxy centre, for all the different runs.
The color palette describes the mass density in each point of that phase space in log $\Msun /  {\rm log \ n /  log \ T}$ units. 
 The distribution of the cold (T$<10^4$ K) and  dense ($n>1 \cmc$) star-forming  gas is smaller with radiation pressure.
 The top-right panel shows the density distribution functions. The different models of radiative feedback control the shape of the high-density tail.
The inclusion of radiation pressure decreases the high-density values of the distribution function by a factor of $\sim$4.
 The addition of infrared trapping decreases the high-density tail by a factor of $\sim$2.}
\label{fig:phaseD}
\end{figure*}

Feedback regulates the density and temperature distribution of the multiphase ISM.
 \Fig{phaseD} shows the phase diagrams of density versus temperature of all gas within 5 kpc for all different runs.
It shows the different gas phases of a multiphase ISM.
The comparison between the NoRadPre and RadPre runs clearly explains why the star formation rate and stellar masses are significantly lower when the effects of radiation pressure are taken into account.
The distribution of cold and dense gas is much smaller with radiation pressure.
The maximum density reached in the ISM decreases significantly with radiation pressure.
This supports the arguments by \citet{Hopkins11}, who argue that radiation pressure mainly regulates the distribution of dense  gas.  

Now we compare the RadPre and RadPre\_IR runs and examine the effects on our simple model of infrared trapping. 
The infrared boost on radiation pressure can quickly disrupt high-density regions. 
The maximum density reached in the RadPre\_IR run is significantly lower than in the RadPre run.
There is however a small effect in the bulk of the star-forming gas, which is $\sim$20\% less massive with infrared trapping. This translates to a 20\% reduction of the stellar content.
However, this effect of infrared radiation could be more relevant in more massive galaxies with higher densities \citep{Hopkins11}.

\Fig{phaseD} also compares the RadPre run with RadPre\_LS.
Unlike radiation pressure, photoheating and photoionization directly change the temperature of the star-forming gas, by heating the gas and by modifying the cooling rates.
Therefore, the RadPre\_LS run yields an increase in the dense gas heated by feedback
to temperatures in the $ 10^4 \ {\rm K} - 10^6 \ {\rm K} $ range.
However, this effect involves little mass, so the star-forming gas only decreases by 4\% and the accumulated stellar mass only decreases by 7\% with respect to the RadPre case.
The effect of heating is not as relevant as radiation pressure for the self-regulation of star formation at these mass scales. 
However, it could be important for much smaller halos, where gas is optically thin to UV radiation, such that radiation pressure has little effect and the gas  can be easily photoevaporated by photoheating  \citep{ThoulWeinberg96, Quinn96, BarkanaLoeb99, Gnedin00, Dijkstra04, Governato07}.
Finally, the RadPre\_LS\_IR run yields the combined effects of IR trapping and local photoheating: the decrease of cold and high-density gas and the increase of warm/hot gas heated by feedback.

The top-right panel of \Fig{phaseD} compares the density distribution function of all runs.
Overall, the distributions are close to log-normal \citep{WadaNorman01}, with a peak around $0.5 \cmc$.
The low-density tail is not affected by radiative feedback.
On the contrary, the different models of radiative feedback control the shape of the high-density tail,
which dominates star formation.
Without radiation pressure, there is a secondary peak at high densities, $n\simeq100 \cmc$.
This up-turn is absent with the inclusion of radiation pressure.
A similar behavior occurs in the simulations of isolated discs by \cite{Hopkins11}.
Most of the gas that accumulates at high densities is dispersed with the inclusion of radiation pressure.
The distribution function at high-density values decreases by a factor of $\sim$4.
Infrared trapping reduces the peak of the distribution, as well as the high-density tail by a factor of $\sim$2, due to the stronger effect of radiation pressure.
On the other hand, the inclusion of photoionization/photoheating shows a similar density distribution than RadPre, with a small up-turn at high densities.
This  excess of mass at high-densities is due to the high-density gas, heated by feedback. The addition of infrared trapping effectively disperses this excess of gas (RadPre\_LS\_IR).

\section{Discussion}
\label{sec:discussion}

The resolution of the simulations reported in this paper is 27 proper pc at $z=3$.
This is better than the 40-pc resolution in \cite{Agertz09b} and much better than the 120-pc of the spline softening length in Eris simulation \citep{Guedes11},
although the SPH kernel could be much smaller.
For the first time, cosmological simulations can resolve 
stellar populations up to Orion-like cluster masses ($\sim 10^3 \ \Msun$) at relatively low redshifts.
This allows more accurate feedback models, in which the unit of mass for stellar feedback is a single low-mass star cluster that contains a handful of massive stars in a very small volume.
 
Radiation pressure depends on the mass and age of the star cluster, as well as the gas column density around it.
After assuming a given IMF,
the parameters of the model are well constrained by theory of stellar evolution and observations of nearby star-forming regions \citep{ThompsonQuataertMurray05}. These include the specific luminosity, the stellar age limit, and the column density threshold.
Furthermore, as demonstrated in Appendix B, we found that the effects of radiation pressure are quite insensitive to variations in these parameters by a multiplicative factor of two. This indicates that the results presented here have a certain level of robustness.

Our model of radiative feedback is purely local, restricted to a single cell and its closest neighbors, so we neglect radiative transfer effects.
The propagation of ionizing radiation to a larger volume of the diffuse ISM \citep{Wise12} is ignored.
 These simulations miss the acceleration of winds by radiation that escapes the galactic disc  
 \citep{MurrayQuataertThompson10,Wise12} or giant clumps \citep{Genzel11,Hopkins12Clumps,DekelKrumholz13}.
Due to the $R^{-2}$ dependence of radiation pressure on distance from a single source, 
these effects are expected to be weaker far from the sources of radiation \citep{KrumholzMatzner09}. 
Therefore,  this local model catches the main effect of radiation pressure in star-forming regions on scales of about 20 pc, the minimum scale resolved by these simulations.
This scale is typical of HII regions around low-mass, star clusters \citep{HH98, Townsley03, Feigelson05, Wang08, Guedel08}. 

Our model of radiative feedback differs from the subgrid models that \cite{Hopkins11} and \cite{Agertz12} implemented in isolated disc simulations. 
In our case, the effect of radiation is expressed in terms of pressure, rather than as direct injection of momentum.
Our approach is more conservative, in the sense that we consider only the ionizing radiation while they include the full stellar luminosity.  In our implementation, the effect of radiation pressure is restricted to the first $\sim$5 Myr of the lifetime of a star cluster. 
Our model also imposes a column density threshold for radiation pressure, below which the ionizing radiation escapes freely. 
In other words, radiation pressure is taken into account only where the material is optically thick to that radiation.
\cite{Hopkins11} and \cite{Agertz12} do not consider such a threshold, as they assume that the typical column densities around star clusters are always high, so that the material is always optically thick. However, if the multiphase nature of the gas is resolved, low-density regions can coexist very close to high-density regions and radiation can diffuse into these optically-thin regions. If no threshold is imposed, the effect of radiation pressure could be overestimated in low-density regions.
%
%
 
These other models of radiative feedback have also neglected the effect of the local UV radiation on the gas cooling and heating due to photoionization and photoheating.
The full self-consistent treatment of photoionization and photoheating is a difficult task beyond the scope of this paper.
Instead, we apply a subgrid model that takes into account the main dependencies of the local UV radiation on the cooling and heating rates.
It uses the results from \textsc{starburst99} and \textsc{cloudy} to model stellar evolution and the radiative transfer effects that occur on scales below the resolution scale.


The inclusion of radiation pressure disperses dense gas in star forming regions of GMC-like scales and masses up to 
$10^7 \ \Msun$.
As a result, these GMC-analogues are quickly disrupted by radiative feedback \citep[see also e.g.][]{Hopkins11}.
However, these GMCs should not be confused with the giant clumps seen in more massive, marginally unstable discs 
\citep{Noguchi99,Elmegreen05a, Bournaud07, Genzel08, DSC, Agertz09b, CDB, Ceverino12, Mandelker13}. 
These giant clumps are more massive ($10^8-10^9 \ \Msun$) and much larger (kpc-scales). 
Therefore, it is much more difficult to disrupt them with star-formation-related processes \citep{KrumholzDekel}.
Instead, these giant clumps can drive a steady wind over many free-fall times \citep{DekelKrumholz13}, as observed in high-z clumpy discs \citep{Genzel11}.
This is apparently at odds with the simulations of  \cite{Hopkins12Clumps} and \cite{Genel12}, where giant clumps are also dispersed after a few internal dynamical times (10-100 Myr), due to the very strong feedback assumed.

For example, \cite{Hopkins12Clumps} use very high values for the infrared optical depth of the clumps, $\tau_{\rm IR}=50-100$, which translates into a factor of $\sim 100$ boost in the momentum injected by the trapped infrared radiation. 
On the other hand, \cite{DekelKrumholz13}, based on simulations by \cite{KrumholzThompson12, KrumholzThompson13},
  argue that a realistic infrared boost is much lower, on the order of unity,
 because this radiation destabilizes the wind and generates large cavities with low optical depth, so radiation can escape easily. \cite{KrumholzThompson12, KrumholzThompson13}
 emphasize that radiation pressure is not always able to drive an outflow in a medium with high column densities of dust, as the amount of momentum injected and the photosphere optical depth are also important.
The model in \cite{Hopkins11}, as well as other phenomenological models of  momentum-driven winds, assumes that the gas could be accelerated to the local escape velocity \citep{MurrayQuataertThompson10}. 
In other words, the outflow solution is imposed, regardless of whether the actual physical conditions for an outflow are fulfilled.

The results reported in this paper are focused on a given halo mass and redshift: $\Mv\simeq10^{11} \ \Msun$ at $z\simeq3$.
This allows us to address the issue of low efficiency of star formation at one important mass scale.
However, it is too risky to extrapolate these results to more massive galaxies or lower redshifts.
This is the first, pilot study of a broader program that includes 35 high-resolution, zoom-in simulations of the formation of galaxies with maximum circular velocities of $V_{\rm c}=100-230 \kms$ at $z\simeq1$. 
Results from the full set will be reported in future work.

\section{Summary}
\label{sec:summary}

%
We investigated the effects of the ionizing photons emitted by massive young stars on the formation of low-mass galaxies in $\Mv\simeq10^{11} \ \Msun$ halos at $z \sim 3$, the main progenitors of local MW-size galaxies.
We modeled radiation pressure from local UV sources in cosmological galaxy formation simulations as a non-thermal pressure that acts in dense and optically thick star-forming regions.
We also included the effect of photoionization and photoheating by local sources on the gas cooling and heating rates. 
The main consequence of the local photoionization is the reduction of the HI peak of the cooling curve.
This prevents cooling in the $10^{4-4.5}$ K temperature range within star-forming regions \citep{GnedinHollon12}.
This suppression of cooling only happens when the conditions of radiation and gas density are such that most hydrogen atoms are ionized, so HI collisional excitations are very rare events.
We also considered a simple model for a modest boosting of radiation pressure due to the trapping of infrared radiation.

%
We find that the main effect of radiation pressure is to regulate and limit the high values of gas density, by 
pressurizing the gas in star-forming regions (\Figs{SFNEW}, \ref{fig:LSNEW}, and \ref{fig:phaseD}).
This prevents the runaway gravitational collapse to high densities. These momentum-driven effects are especially effective in dense regions, which radiate thermal energy very efficiently and thus make the thermal feedback ineffective in reducing the local high gas density.

By suppressing the high densities, radiation pressure regulates star formation.
The star-formation history is generally a factor 3 lower when radiation pressure, as implemented here, is taken into account (\Fig{SFR}).
We find that the typical star-formation rate of a MW progenitor at $z=3$, within a $\Mv\simeq10^{11} \ \Msun$ halo, is about $1 \ \Msun {\rm yr}^{-1}$.
This is roughly consistent with the average values coming from abundance matching models \citep{Behroozi12}. 
In our simulations, radiation pressure from local ionizing sources drives the low efficiency of galaxy formation in 
these low-mass halos.
The stellar-to-virial mass ratio is as low as f$_{\rm stars}\simeq 0.6\%$ (\tab{3}).
This value is consistent with current abundance matching models \citep{Moster12, Behroozi12}, within the systematic uncertainties.
The effect of infrared trapping yields 20\% changes in stellar mass, so it is a secondary effect.
Finally, photoionization and photoheating modify the stellar distribution by 10\%.

We find that radiation pressure generates a more extended stellar distribution.
As opposed to a compact and round stellar core, the runs with radiation pressure show a prolate stellar distribution. 
Although the gas is distributed in a turbulent and irregular disc, only a dense, bar-shaped region can form stars.
Radiation pressure impedes the formation of stars everywhere in the disc and it keeps a high gas fraction, 
$ M_{\rm gas} / \Ms  \geq 1$ within the galaxy.
The more extended and less massive stellar distribution yields a rising circular velocity profile (\Fig{Vc}), 
because
baryons only dominate the mass in the inner 0.3 kpc and the dark matter contribution to the circular velocity gently rises toward larger radii.

\section*{Acknowledgments} 
 
We acknowledge discussions with Rachel Sommerville, and Risa Wechsler.
The simulations were performed at NASA Advanced Supercomputing (NAS) at NASA Ames Research Center.
This work was partially supported by MINECO  (Spain) - AYA2012-31101 and MICINN (Spain)  AYA-2009-13875-C03-02.
DC is a Juan de la Cierva fellow.  AK, EK, and CWC were partially supported by Program number  AR-12646,
provided by NASA through a grant from the Space Telescope Science
Institute, which is operated by the Association of Universities for
Research in Astronomy, Incorporated, under NASA contract NAS5-26555.
AD was supported by ISF grant 24/12,
by GIF grant G-1052-104.7/2009,
by a DIP grant,
by NSF grant AST-1010033,
and by the I-CORE Program of the PBC and the ISF grant 1829/12.

\bibliographystyle{mn2e}
\bibliography{Rad_Feedback5}

\appendix

\section{A model of star formation for scales between 10-30 pc}

\begin{figure} 
\includegraphics[width =0.49 \textwidth]{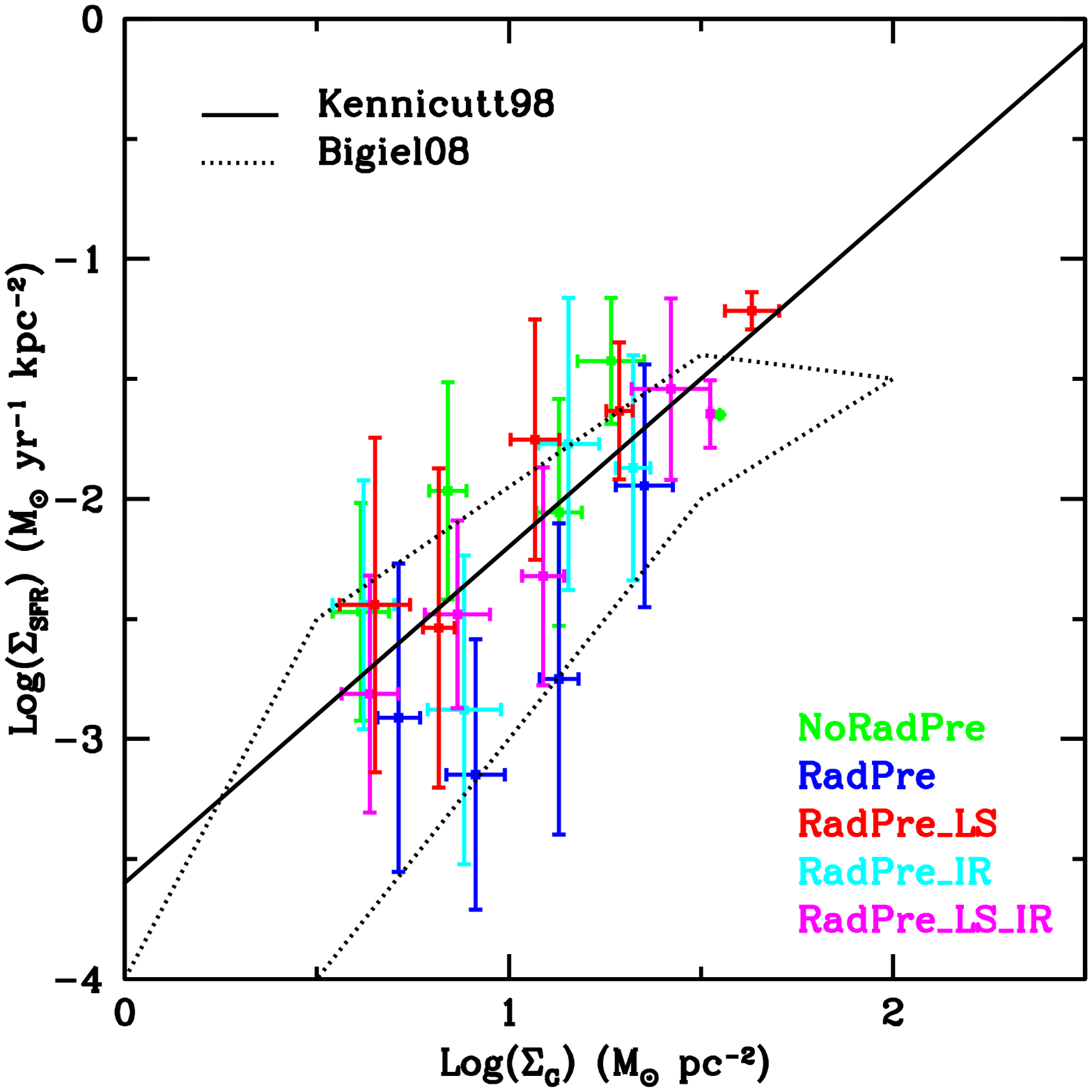}
\caption{The Kennicutt-Schmidt relation for the different runs. Each point represents  the median of the star-formation surface density at a given bin of gas surface density. The error bars account for $\pm 1 \sigma$ dispersion of the values within the bin. 
The black solid  line marks the relation by Kennicutt et al. (1998) and the dotted lines contour the 
distribution of sub-kpc-sized patches in the sample of nearby galaxies by Bigiel et al. (2008). All runs are roughly in agreement with these observed relations.}
\label{fig:kenn}
\end{figure}

The model for star formation is based on the model first described in the appendix of CK09.
Star formation is allowed in a time step, $dt_{\rm SF}=5$ Myr.
 During this period of time, a stellar particle can be formed only where the density 
reaches a given threshold: $\rho_{\rm gas} > \rho_{\rm SF}$ and the temperature is below a given value,  $T_{\rm gas} < T_{\rm SF}.$ 
Even in these cold and dense regions,
each star formation event is treated as a random event with a
probability $Pr$ to occur. We roughly approximate the fact that
regions with higher densities have shorter free-fall times and they have higher probabilities to host star
formation events by assuming a simplified formula:
\begin{equation}
Pr  = {\rm min} \left (0.2, \sqrt{ \frac{ \rho_{\rm gas}} { 1000\rho_{\rm SF}}} \right ) 
\end{equation}
In this way, the number of stellar particles remains in a range that
is not computational prohibited.
In the formation of a single stellar particle, the star formation rate
is proportional to the gas density \citep{Kravtsov03}:
\begin{eqnarray}
\frac{ d \rho_{*, \rm young}}{dt} & = & \frac{\rho_{\rm gas}}{\tau}
\label{eq:sf}
\end{eqnarray}
where $\rho_{*, \rm young}$ is the density of new stars, $\rho_{\rm
gas}$ is the gas density and $\tau$ is a constant star formation
timescale.  The density and temperature thresholds used are $\rho_{\rm
SF}=0.035$ $\Msun$ pc$^{-3}$ $(n=1 $ $\cmc)$ and $T_{\rm SF}=10^4 $
K. In spite of the fact that we allow star formation starting at $T<10^4
$ K, in practice the vast majority ($>90\%$) of ``stars'' form at
temperatures below 1000~K and more than half of them form around
300 K.
Equation \ref{eq:sf} can be integrated over a single cell and over  $dt_{\rm SF}$ 
and it yields:
\begin{eqnarray}
\frac{m_{*,\rm young}}{m_{\rm gas}} & = & \frac{dt_{\rm SF}}{\tau}
\end{eqnarray}
where $m_{*,\rm young}$ is the stellar mass of the new stellar particle and $m_{\rm gas}$ is the mass of gas in that cell.
The value of $\tau=12$ Myr was calibrated in order to reproduce the empirical Kennicutt-Schmidt law  \citep{Kennicutt98} for the resolution of these simulations (\Fig{kenn}).
Each point in the figure corresponds to the median of the star-formation surface density at a given bin of gas surface density.
Both surface densities are computed within different 1 kpc$^{2}$ square patches in the face-on view of the gaseous disc.
For the computation of the star-formation surface density, stars younger than 100 Myr are used.
All runs are roughly in agreement with \cite{Kennicutt98} results
and with the distribution of sub-kpc sized patches in the sample of nearby galaxies by Bigiel et al. (2008).
The NoRadPre run slightly overproduces stars and the RadPre run slightly underproduces them, although the scatter between different patches in the same bin is large.

\section{Robustness of results to variation of the parameters of the radiation pressure model and convergence}

\begin{figure} 
\includegraphics[width =0.49 \textwidth]{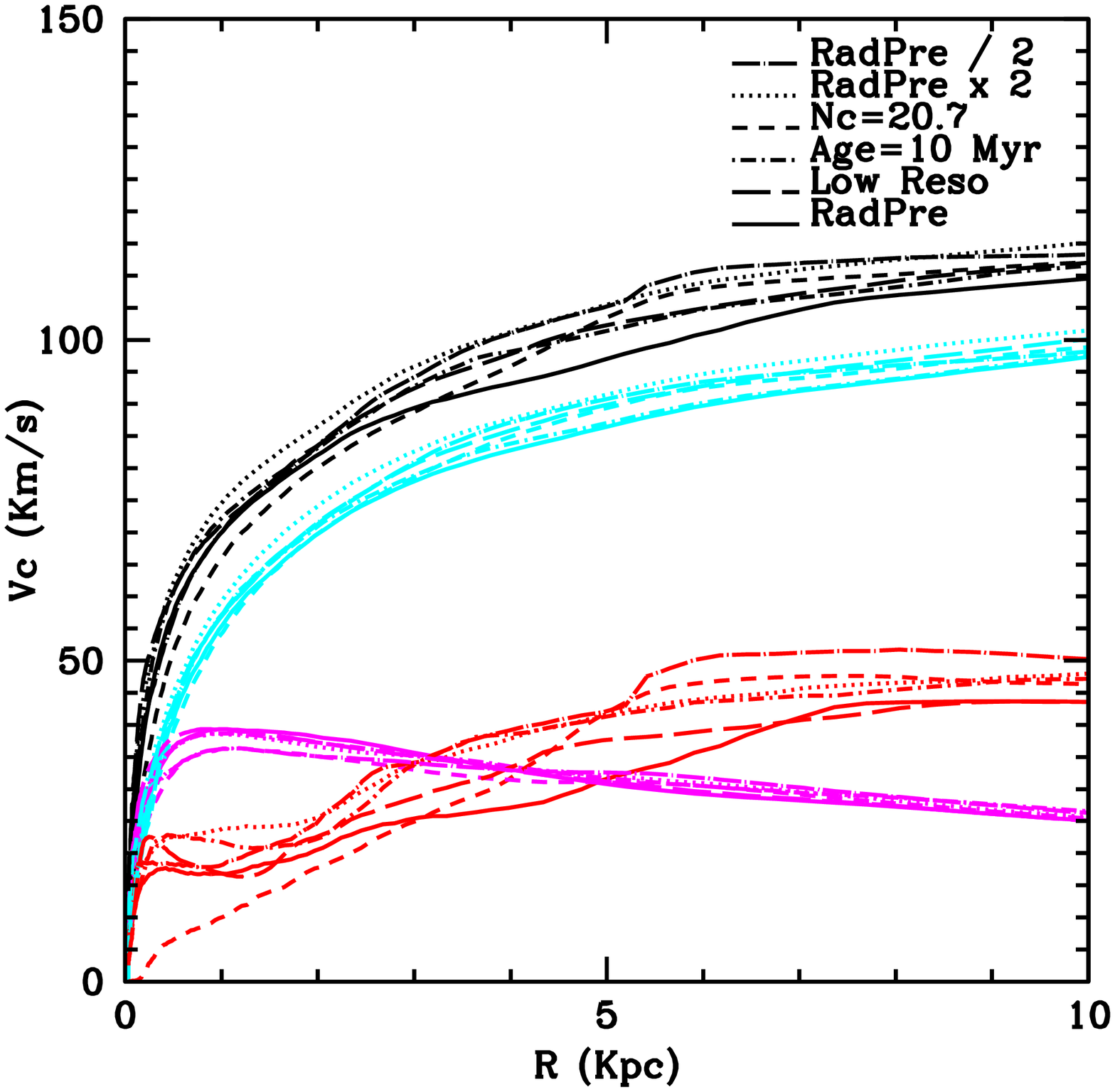}
\caption{Circular velocity profile, analogous to \Fig{Vc}, but for different variations in the parameters of the model for radiation pressure plus a low-resolution run of the fiducial RadPre model. All runs show very similar results.}
\label{fig:var}
\end{figure}

\Fig{var} shows the circular velocity profile, similar to \Fig{Vc}, but for different variations in the parameters of the model of radiation pressure (\se{RadPre}).
In particular, we consider 5 different variations: a run that doubles the specific luminosity, $\Gamma'= 2 \times 10^{36}$ erg s$^{-1} \msun^{-1}$; 
a run with half of the fiducial specific luminosity, $\Gamma'= 5 \times 10^{35}$ erg s$^{-1} \msun^{-1}$;
a run in which the column density threshold is half the fiducial value, $N_{\rm rad}= 5 \times 10^{20} \cms$; and
a run in which the period of radiation pressure has been doubled, t$_{\rm age}$=10 Myr.
All runs show very similar results. 
Therefore, our conclusions about the effects of radiation pressure are robust against variations of a factor of 2-4 in the parameters of the radiation pressure model.

\Fig{var} also shows the circular velocity profile for a twice lower resolution (54 proper pc at $z=3$) run of the RadPre model.
The circular velocity profile shows good convergence. The differences between the high-resolution and low-resolution runs are similar to the differences between runs with different values of the parameters of the model of radiation pressure. The stellar mass differs only by 2\%. The biggest difference occurs in the gas. At a radius of 4 kpc, the gas mass is 33\% higher than in the high-resolution run.

\bsp

\label{lastpage}

\end{document}